\documentstyle[aps,epsf]{revtex} 

\begin{document} \renewcommand{\thefootnote}{\fnsymbol{footnote}}

\draft  
\hsize\textwidth\columnwidth\hsize\csname@twocolumnfalse\endcsname\title {Two
band gap field-dependent thermal conductivity of $MgB_2$}

\author{Bai-Mei Wu$^{1,2}$\footnotemark[1], Dong-Sheng Yang$^{1}$, Wei-Hua
Zeng$^{1}$, Shi-Yan Li$^{1}$, Bo Li$^{1}$, Rong Fan$^{1}$, \\Xian-Hui Chen$^{1}$,
Lie-Zhao Cao$^{1}$ \\and Marcel Ausloos$^2$\footnotemark[2]}

\address{$^1$ Structural Research Lab., Dept. of Physics, University of Science
and Technology of China, Hefei, Anhui 230026\\ $^2$SUPRAS Institut de Physique
B5, Universit\'{e} de Li\`{e}ge, B-4000 Li\`{e}ge, Euroland}

\date{\today} \maketitle

\begin{abstract}

The thermal conductivity $\kappa (H,T)$ of the  new superconductor $MgB_2$ was
studied as a function of the temperature and a magnetic field.  No anomaly in
the thermal conductivity $\kappa (H,T)$ is observed around the superconducting
transition in absence or presence of magnetic fields up to 14 Tesla;  upon that
field the superconductivity of $MgB_2$ persisted.  The thermal conductivity in
zero-field shows a $T$-linear increase up to 50K.  The thermal conductivity is
found to increase with increasing field at high fields. We interpret the findings
as if there are two subsystems of quasiparticles with different field-dependent
characters in a two ($L$ and $S$)-band superconductor reacting differently with
the vortex structure. The unusual enhancement of $\kappa (H ,T)$ at low
temperature but higher than a ($H_{c2S}\simeq 3T$) critical field is interpreted as a result of
the overlap of the low energy states outside the vortex cores in the $S$-band.
\end{abstract}

\footnotetext[1]{electronic address: wubm@ustc.edu.cn}
\footnotetext[2]{electronic address: marcel.ausloos@ulg.ac.be}

\section{INTRODUCTION}

Superconductivity at a remarkably high transition temperature closed to 40K was
recently discovered in $MgB_2$ [1,2].  $MgB_2$ has a Debye temperature about
800K, several times larger than that of $Nb_3Sn$, as seen from specific heat
measurements [3-7]. The phonon density of states of $MgB_2$ has been obtained by
inelastic neutron scattering [8-11].  These results indicate that phonons are
playing an important role in the interaction of the new superconductor $MgB_2$.
Most experiments in $MgB_2$, such as the isotope effect [12,13],  $T_c$ pressure
dependence [14,15] and tunneling spectroscopy [16,17] indicate that the
superconductivity of $MgB_2$ can be consistent with a phonon-mediated BCS
electron pairing.  Beside [5-7,17] reports have also shown some evidence for two
gaps in the quasiparticle excitation spectrum of  $MgB_2$ [18-20].

The thermal conductivity has been used widely to study superconductors [21],
offering important clues about the nature of heat carriers and scattering
processes between them, especially in $H$=0, and in unconventional
superconductors [22-26], the more so in the superconducting state for which
traditional electrical probes such as the electrical resistivity, Hall effect,
and thermopower are inoperative. The study of the thermal conductivity in applied
magnetic fields has also been an interesting new probe of the vortex state of
high- $T_c$ superconductors [27-37] and in $MgB_2$, in absence [38-40] or
presence [41-42] of a magnetic field.

In this paper we present a new study of the thermal conductivity $\kappa (H,T)$
of polycrystalline $MgB_2$, as a function of temperature in both absence and
presence of magnetic fields in order to probe the anomalous dependences [40,41]
in light of new vortex structure considerations [7].  The temperature range goes
between 4-50K and the applied magnetic field goes up to 13.9 Tesla. The
superconductivity of $MgB_2$ persisted over the investigated condition though no
sharp  anomaly in the thermal conductivity $\kappa (H,T) $ was observed around
the superconducting transition.  The thermal conductivity almost monotonously
increases with increasing temperature in zero and  low magnetic fields whereas a
significant enhancement of the values of the thermal conductivity is found for
high magnetic fields.  A model based on two different gaps and two subsystems of
quasiparticles with different field-dependent characters in superconducting
$MgB_2$ is argued to be consistent with these results.

\section{SAMPLE PREPARATION AND MEASUREMENTS}

Polycrystalline $MgB_2$ samples  were prepared by the conventional solid state
reaction method.  A mixed powder of high purity $Mg$ and $B$ with an appropriate
ratio was ground and pressed into pellets.  The pellets were wrapped in a
tantalum foil and sealed in a stainless steel reactor, then heated at 950$^\circ$
C for 4 hours under an argon gas flow.  Repeating the process we got a
high-density bulk sample.  The phase purity was confirmed by X-ray diffraction
analysis.  No impurity phase was observed within the usual X-ray resolution.  For
the electrical and thermal conductivity measurements one 8 x 1.5 x 0.8 mm$^3$ bar
was cut from a bulk sample pellet.

In order to check the sample superconductivity we have measured the electrical
resistivity by the standard four-probe d.c. technique in the 4 to 50 K
temperature range and for magnetic fields up to 14 T. Fig. 1 shows the
temperature dependence of the electrical resistivity, $\rho (H ,T) $, of
polycrystalline $MgB_2$ for several magnetic fields under heating condition. The
applied field was 0, 1, 4, 7, 10 and 14T in turn applied at low temperature after
zero field cooling. A narrow superconductivity transition occurs at $T_c$ = 38.1K
in zero-field.  With increasing fields $T_c$ shifts to lower temperature and the
transition width increases as expected.  It should be noted that the
superconductivity of $MgB_2$ persisted over the investigated fields.

The thermal conductivity of the sample was measured by the longitudinal
steady-state thermal flow method in the temperature range 4-50 K and for several
magnetic field up to 13.9 T. The thermal flow and the magnetic field were
parallel to the long axis of the sample.  A thin-film chip resistor was used as
an end heater, and thin-wire differential thermocouples for monitoring the
thermal gradients on the sample. The measurements were performed under high
vacuum with two additional shields mounted around the sample in order to reduce
the heat losses due to radiation at finite temperature [43]. A carbon-doped glass
resistance thermometer was used to measure the temperature of sample in a
magnetic field.  The sensitivity of the thermocouple as well as the resistance of
the thin-film resistor were previously calibrated for magnetic field conditions.
The whole experiment was computer controlled [44].

Fig. 2 shows the temperature dependence of the thermal conductivity of
polycrystalline $MgB_2$, in applied  0.0, 0.02, 0.5, 1.0, 5.0, 10.0 and 13.9
Tesla field after zero-field cooling.  Data were recorded while warming the
sample up at a fixed magnetic field.  The thermal conductivity is seen to
increase monotonically with increasing temperature in the investigated ranges.
The temperature dependence of $\kappa (0, T)$ is linear (correlation coefficient
R = 0.9995).  With increasing fields, $\kappa ( H , T)$  has an almost linear
temperature dependence at low fields.  A significant enhancement in $\kappa
(H,T)$ is obvious in the high- field region at low temperature, as in [41,42].
All $\kappa ( H , T)$  over $T$ curve intersect each other at 19K above which
temperature the curves of $\kappa (H ,T)$ for different fields merge on each
other, except that for 13.9T.

\section{DISCUSSION of MEASUREMENTS}

As shown in Fig.2, the absence of anomaly in the thermal conductivity in
zero-field   near the superconducting transition $T_c$, is first in reasonable
agreement with other results reported for polycrystalline [37-39] and single
crystalline $MgB_2$ [41,42]. Such a smoothness is often attributed to charge
carrier and phonon scattering by a distribution of localized impurities or
extended defects.  However, the $MgB_2$ thermal conductivity markedly differs in
behavior at $T_c$ from that of conventional superconductors and high temperature
superconductors.  The thermal conductivity of such superconducting materials
generally shows an anomalous behavior at the superconducting transition
temperature, either a decrease or an increase below $T_c$ depending on the nature
of the heat carrier and their interaction.  In conventional  superconductors
$\kappa (T)$ decreases below $T_c$ due to the condensation of electrons.  In the
case of the high temperature superconductors $\kappa (T)$ shows a peak below
$T_c$ that is explained by considering either an increase in the phonon mean free
path due to carrier condensation or an increase in the quasi-particles relaxation
time in the superconducting state or to both.  Our results show no such a
response in the thermal conductivity for $MgB_2$ around $T_c$ in both absence
(and presence of magnetic fields) within our experimental resolution.  It means
that $MgB_2$ thermal conduction is not much sensitive to any Cooper pairing below
$T_c$ nor to the breaking effect introduced by a field. However, the lattice
contribution is substantial in both the superconducting and the normal state
region. Indeed, in zero field, the thermal conductivity $\kappa$ in a metal is
generally considered to be the sum of electrons $\kappa_e$  and phonon
contributions $\kappa_{ph}$, i.e. $\kappa (T) $  = $\kappa_e (T) $  +
$\kappa_{ph} (T) $.

The electronic contribution to $\kappa(T)$ in the superconducting state has been
calculated by Bardeen, Rickayzen, and Tewordt [21]. The phonon thermal
conductivity is usually represented within the Debye-type relaxation rate
approximation [45] \begin{equation}\label{eKph} \kappa_{\rm ph} =  \frac{ k_{\rm
B}}{ 2 {\pi}^{2} v } \left( \frac{k_{\rm     B}}{\hbar} \right) ^{3} T^{3}
\int\limits_{0}^{\Theta_{D}/T} \frac{x^{4}e^{x}}{(e^{x}-1)^{2}}\tau (\omega,T)
dx, \end{equation} where $\omega$ is the frequency of a phonon, $\tau(\omega,T)$
is the corresponding relaxation time, $ v=\Theta _D\left( {{{k_B} \mathord{\left/
{\vphantom {{k_B} \hbar }} \right. \kern-\nulldelimiterspace} \hbar }}
\right)(6\pi ^2n)^{-1/3}$ is the average sound velocity,  $n$ is the number
density of atoms, and $x=\hbar\omega/k_{\rm B} T$. For our analysis, we use
$\Theta_{D}=750$~K, as deduced from specific heat measurements [4,9]. The total
phonon relaxation rate \begin{eqnarray}\label{eTau} \tau^{-1} = L/v +   B
\omega^{4} + C T \omega^{2} \exp(-\Theta_{D}/bT) + D \omega\nonumber\\ + E \omega
g(x,y) \end{eqnarray} can be represented as a sum of terms  corresponding to
independent scattering mechanisms. The individual terms, dominating in different
temperature intervals, introduce phonon scattering by sample boundaries, point
defects, phonons, dislocations, and electrons, respectively. The constants $L$,
$B$, $C$,  $b$, $D$, and $E$ are a measure for the intensity of corresponding
phonon relaxation processes. The function $g(x,y)$ is given in Ref.[21].

The $\kappa_e (T) $ magnitude can be estimated from the electrical resistivity
and the Wiedemann-Franz (W-F) rule $\kappa_e \rho = LT$ where $L$ = $L_0$, the
Lorentz  constant, if the scattering is elastic; $L<L_0$ if it is inelastic.
Since the value of $\kappa_e$  is an upper limit when obtained by using the W-F
rule and $L_0$, we estimate that the value of $\kappa_{e}$ is about 20\% of the
$total$ value of the thermal conductivity in the normal state, $\kappa_{n}$ e.g. at $T$=50K
for our sample. The best fit value for $L$ being significantly smaller than the
expected Lorentz constant $L_0$ calls attention to the influence of inelastic
scattering, and to a large phonon (80\%) contribution.

Next, consider the field dependence of the thermal conductivity $\kappa
(H)$/$\kappa (H=0)$ at several temperatures (Fig.3). In low fields, the thermal
conductivity seems insensitive to the magnitude of the applied magnetic fields
whereas it becomes variably enhanced at higher fields.  Notably, the value
$\kappa (H)$ is higher than that of $\kappa (H=0)$ in the field region larger
than 0.5 T at low temperature (below 19K). The non-linear field dependence and
the field-dependence enhancement in $\kappa (H)$ imply that the effect of vortex
structure, induced by the field, over heat carrier in the mixed state of $MgB_2$
is no longer a beaten track.  According to standard thinking for conventional
superconductors the introduction of a magnetic field would lead to a decrease in
heat conductivity because vortices constitute new scattering centers for heat
carriers.  This picture provided a qualitative explanation of the field-induced
decrease in $\kappa (H)$ observed in the vortex states of conventional
superconductors and high $T_c$  cuprates in the investigated regions of the
field-temperature plane [27-37].

To interpret quantitatively the field dependence of the thermal conductivity in
superconductors one again assumes that phonons and electrons contribute
independently as thermal carriers, such that \begin{equation} \kappa (B,T)  =
\kappa_e (B,T)  + \kappa_{ph} (B,T) , \end{equation} where $\kappa_e (B,T)$ and
$\kappa_{ph} (B,T)$ are the electronic and phononic contribution to $\kappa (B,T)$,
respectively.

Again the mean free path of both, electrons and phonons, can change below $T_c$,
because the scattering of phonons on electrons is reduced when electrons condense
into the superconducting state. Moreover the electron-electron interaction can
change when the superconducting gap opens below $T_c$ and modify the mobility
because of the condensate presence. Therefore, it is not immediately clear {\it a
priori} which type of thermal carriers and which interactions are responsible for
the behavior of the thermal conductivity in various temperature and field ranges.

More recently experiments have shown unequivocally that the two gaps structure is
an intrinsic Fermi property of $MgB_2$.  According to band structure calculations
and tunneling experiments, there are two distinctive Fermi surfaces : one is a
two dimensional cylindrical Fermi surface arising from $\sigma $-orbitals due to
$p_x$ and $p_y$ electrons of $B$ atoms and the other is a three dimensional
tubular Fermi surface network coming from $\pi$-orbitals due to $p_z$ electrons
of $B$ atoms. They are weakly hybridized with $Mg$ electron orbitals. These two
Fermi surfaces have different superconducting energy gaps: A large band gap (LBG)
$\Delta_L$  on the 2D Fermi surface sheets and a small band gap (SBG) $\Delta_S$
on the 3D Fermi surface. The ratio $\Delta_S$ /$\Delta_L$  is estimated to be
around 0.3$-$0.4 [7,17,41].  Electrons on these two Fermi surfaces own different
characters since the $\sigma$-orbital is strongly coupled to the in-plane
$B$-atom vibration with $E_{2g}$ symmetry but the $\pi$-orbital is weakly coupled
with this phonon mode

Consider first the possible electronic contribution and let us analyze the field
dependence of the thermal conductivity at low temperature (19K in our study)
within a vortex lattice structure picture for a two band gap superconductor. The
single vortex state has been studied by Nakai et al. [7]. In each unit cell and
within a two-band superconducting model [7], the vortex core radius is found to
be narrow for the LBG and is large for the SBG electrons.  When increasing $H$
the vortex core radius widens (and the order parameter is suppressed).  This
widening $H$ effect is stronger for the SBGÊ than for the LBG  electrons.
Therefore the LBG vortex bound states are highly confined, due the narrow core
radius, while the S-band vortex core states are more loosely bound.  In the
vortex core of the L-band, the local density of states (LDOS) at site $r$, $N_L
(r, E \sim O)$, is thus highly concentrated, while the $N_S (r, E \sim O)$ LDOS
is rather spread out in the $S$-band case.  In fact, the low energy states
extending from the $S$-band vortex cores are even expected to overlap with
neighboring ones. For increasing $H $the overlap is expected to become more
pronounced and the LDOS to reduce to a flat profile, i.e. $N_S (r, E \sim O)/
N_s(E_F) \simeq 0.5$ for the $S$-band, - where $N_S (E_F)$ is the total DOS in
the normal state at the Fermi level. This picture of the vortex structure in
$MgB_2$ is in fine agreement with the field dependence of the electronic specific
heat [5-7].

From a scattering point of view,  we have also to consider two subsystems of
quasiparticles  characterized by $\Delta_L$ and $\Delta_S$ band gaps.  Let
$\kappa _e (B, T) = \kappa_{e,L}(B, T) + \kappa_{e, S}(B, T)$, thereby taking
into account two sets of electrons with different characters.

According to the analysis on single crystalline $MgB_2$ [41] the electrons
experiencing the large gap provide approximately 2/3 of the total electronic heat
conduction, and the dominant part of the interaction between quasiparticles and
low-frequency phonons is provided by that part of the electronic excitation
spectrum experiencing the small gap [25,41].

The function $g(x,y)$ in Eq.(2) is then given by the sum
\begin{equation}\label{eEP2Gaps} g(x,y) =  g(x,\Delta_L(T)/k_B
T)  +  \alpha g(x,\Delta_{S}(T)/k_B T). \end{equation} The parameter
$\alpha$ characterizes the relative weight of phonon scattering by quasiparticles
in the two $S$ and $L$ subsystems.

It seems reasonable to  expect that a weak field more easily suppresses the
energy SBG than the LBG. Therefore superconductivity is maintained in a field
mainly because the LBG survives, up to $H_{c2}$, as seen in Fig.1, i.e $\alpha =
0$. The $ N_s (r, E \sim O)$ domains with low-energy excitations survive, but for
increasing fields the SBG LDOS reaches a flat profile $N_s (r, E \sim O) /
N_s(E_F)$ $\simeq0.5$, thereby not contributing to heat conduction.   This
$H_{c2S}$ seems to be ca. 3.0 T.  These SBG electrons become heat carrier prone,
inducing above $H_{c2S}$ an increase in the electronic thermal conductivity.
Nakai et al. [7] claim that $H_{c2S}$ and $H_{c2L}$ are the same because they do
not observe a hump in the specific heat at some intermediary field, i.e. the
lower one. This maybe due to the fact that the specific heat is on one hand a
bulk quantity in contrast to the thermal conductivity rather probing a
percolation path. Moreover, we have shown elsewhere that it is sometimes very
hard to obtain accurately kinks and phase transtion lines form specific heat data
[46], even with high precision measurements[47], in contrast to magnetotransport
data [48].

Consider the phonon contribution next. Phonons are accepted to be the dominant
heat carriers in $MgB_2$, especially down to  $T$/ $T_c$ $\simeq$ 0.4 [37-39].
The study of the phonon spectrum in $MgB_2$  [8-11] has clearly shown an
anomalous phonon behavior at the low energy transfer of -24 meV on cooling
through $T_c$ reflecting a strong relation between these phonon states and
superconductivity origin in $MgB_2$.

From  the thermal conductivity  in a magnetic field behavior point of view, the
phonon-(double type) vortex structure interaction should be expected to depend
also on the nature of the phonon spectrum, in the sense of Tewordt-Wolkhausen
picture [22]. The phonons which can inelastically interact with the bound
quasiparticles in the vortex cores are only those for which the phonon wavelength
$\lambda <  \xi$, is of the size of a vortex core [27].  Thus it seems obvious
that due to the overlapping features of the low energy states, the SBG vortices
become less effective scattering centers for phonons as well.  Therefore the mean
free path of low frequency phonons may also increase, resulting in an increase of
the thermal conduction process at high field, in agreement with  the above data.
We conjecture that the quasi particle associated with the LBG contribute
moderately to the thermal conductivity at high field and low temperature. In fact
interactions involving the quasiparticles associated with the LBG and thermal
phonons have not been well described. Nevertheless they could contain an elastic
component of the scattering process. Such a picture seems a fine way of
interpreting the decrease of the thermal conductivity at high field and high
temperature (Fig.2) .

Finally,  an alternative view considers  that electrons self-organize into
collective textures rather than point-like entities [49].  Such an image of
electronic collective textures in an applied magnetic field is complex and not
immediately ready for interpreting the (electron or phonon in fact) scattering
process within the framework of  a double vortex structure, as discussed above.
We cannot further debar the possibility that a  field-induced shift of the phonon
spectrum itself could produce effects similar to temperature.  Some softening of
the phonon spectrum indeed exists, as seen in the thermal expansion coefficient
[50,51]. This equivalency, to be fair, may be also at the origin of the enhanced
behavior of the field-temperature dependent thermal conductivity in
superconducting $MgB_2$.

\section{CONCLUSION}

In conclusion, we have investigated the heat transport in $MgB_2$ in both absence
and presence of magnetic fields (up to 14T) in the temperature range between
4-50K. In these regimes, the superconductivity of $MgB_2$ persisted as seen from
electrical resistivity measurements; no clear anomaly appeared in the thermal
conductivity at $T_c$. The thermal conductivity in absence of field shows a
$T$-linear increase up to 50 K. In weak magnetic fields the thermal conductivity
is insensitive to the magnitude of the applied field; an enhanced thermal
conductivity appears at higher fields and low temperature, but not at high
temperature. We argue that the existence of two quasiparticle subsystems with
different field-dependent characters explain the results. The field dependence of
the thermal conductivity for such a two-band superconductor indicates that the
enhancement of $\kappa(H,T)$ at low temperature in high fields is a result of the
overlap of the low energy states outside the small band gap vortex cores, above a critical $H_{c2S}$ field.

\vskip 0.5cm {\bf Acknowledgements}

This work was supported by the National Natural Science Foundation of China
(No.10174070), the Ministry of Science and Technology of China (NKBRSF-G19990646)
and partly supported by the Belgium National Fund for Scientific Research (FNRS)

\vskip 0.3cm

\section*{Figure captions} \vspace*{0.2cm} Figure 1:\\ Electrical resistivity of
polycrystalline $MgB_2$  versus temperature in 0,  1, 4, 7, 10 and
14 Tesla applied magnetic fields

\vskip 0.3cm

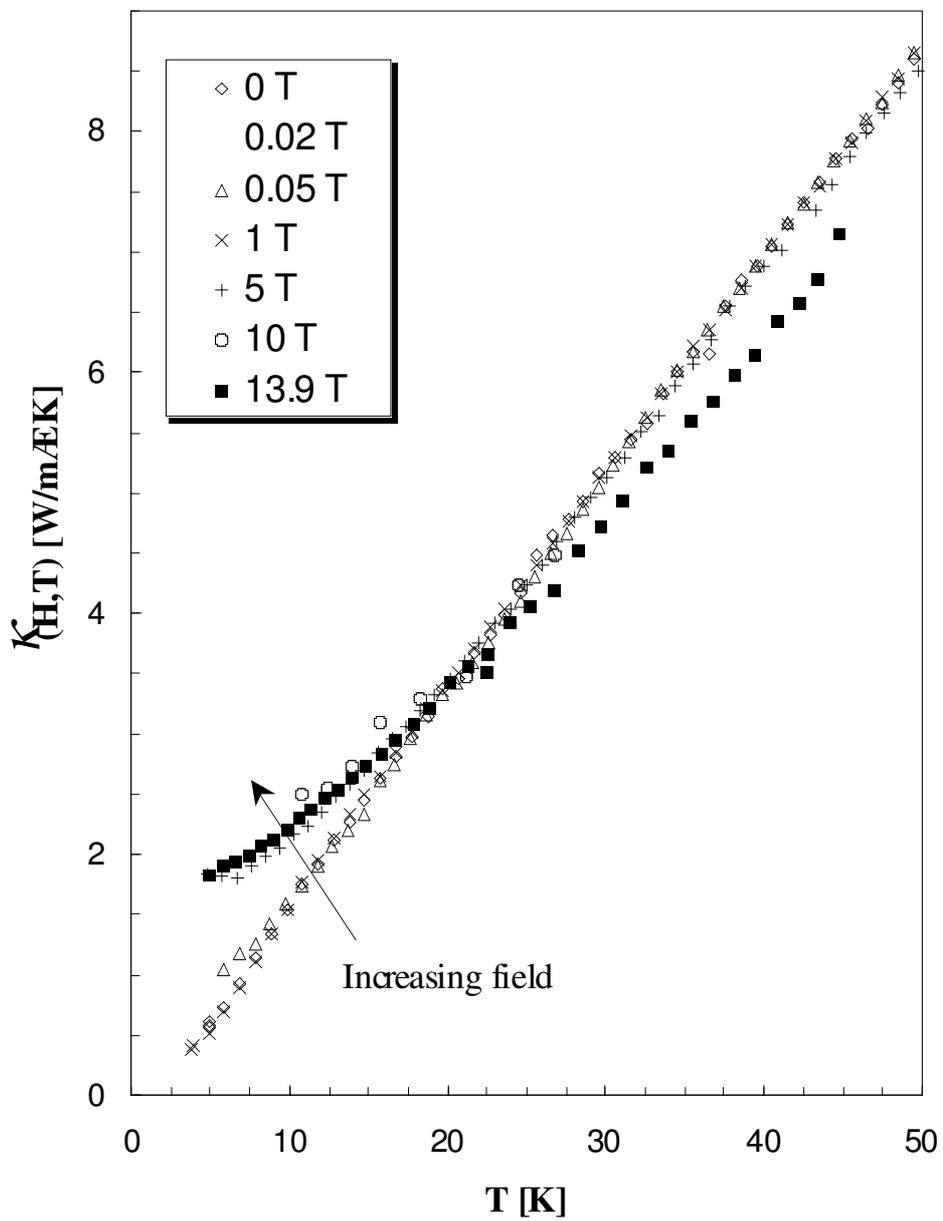
Figure 2:\\ Thermal conductivity, $\kappa(H, T)$, of polycrystalline $MgB_2$ vs.
temperature $T$ in zero $H$-field and in  0.02, 0.5, 1.0, 5.0, 10.0 and 13.9 Tesla
applied magnetic fields

\vskip 0.3cm

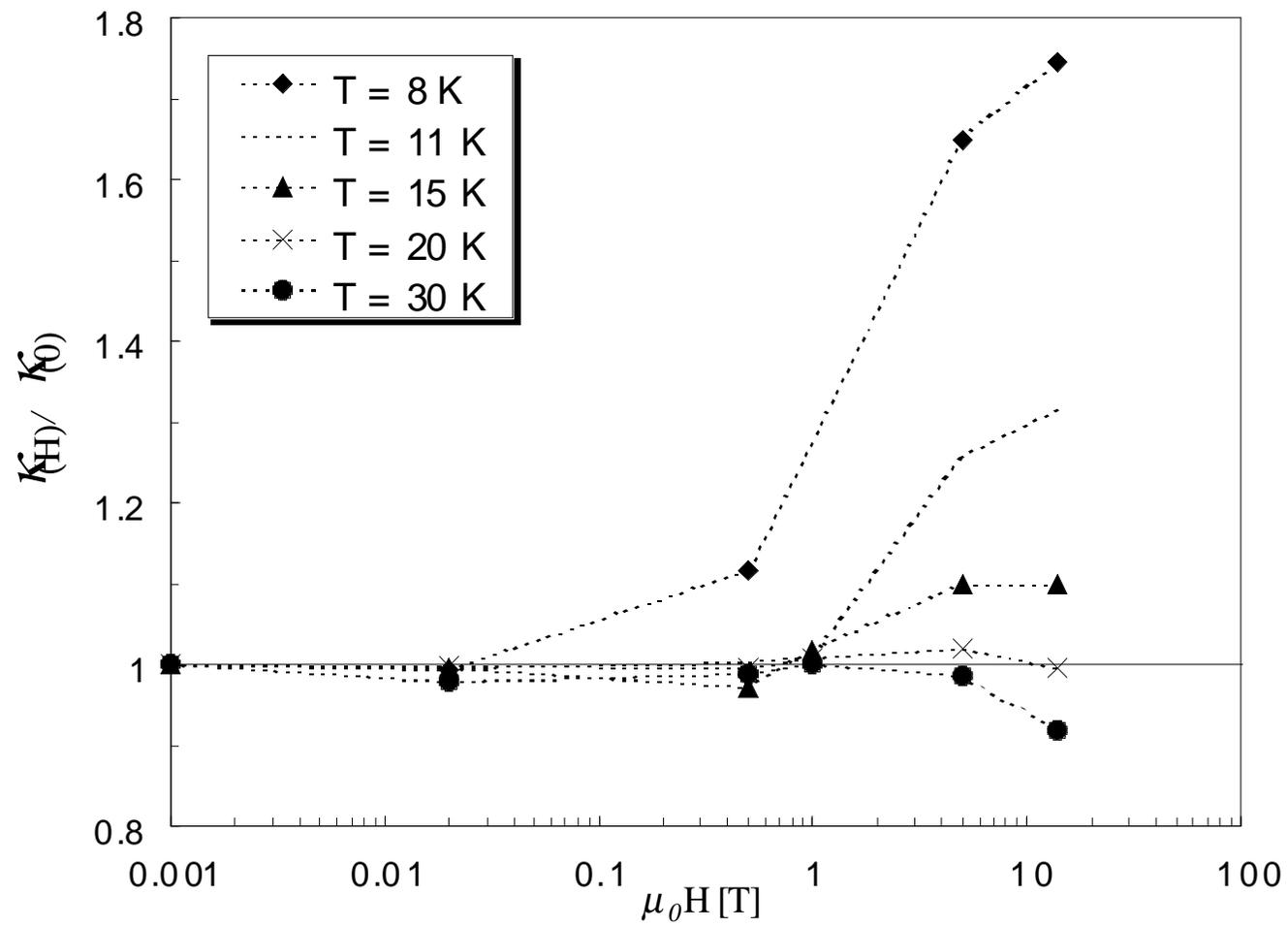
Figure 3:\\ Field dependence of  $MgB_2$ reduced thermal conductivity $\kappa(
H)$/$ \kappa(H=0)$) at several temperatures

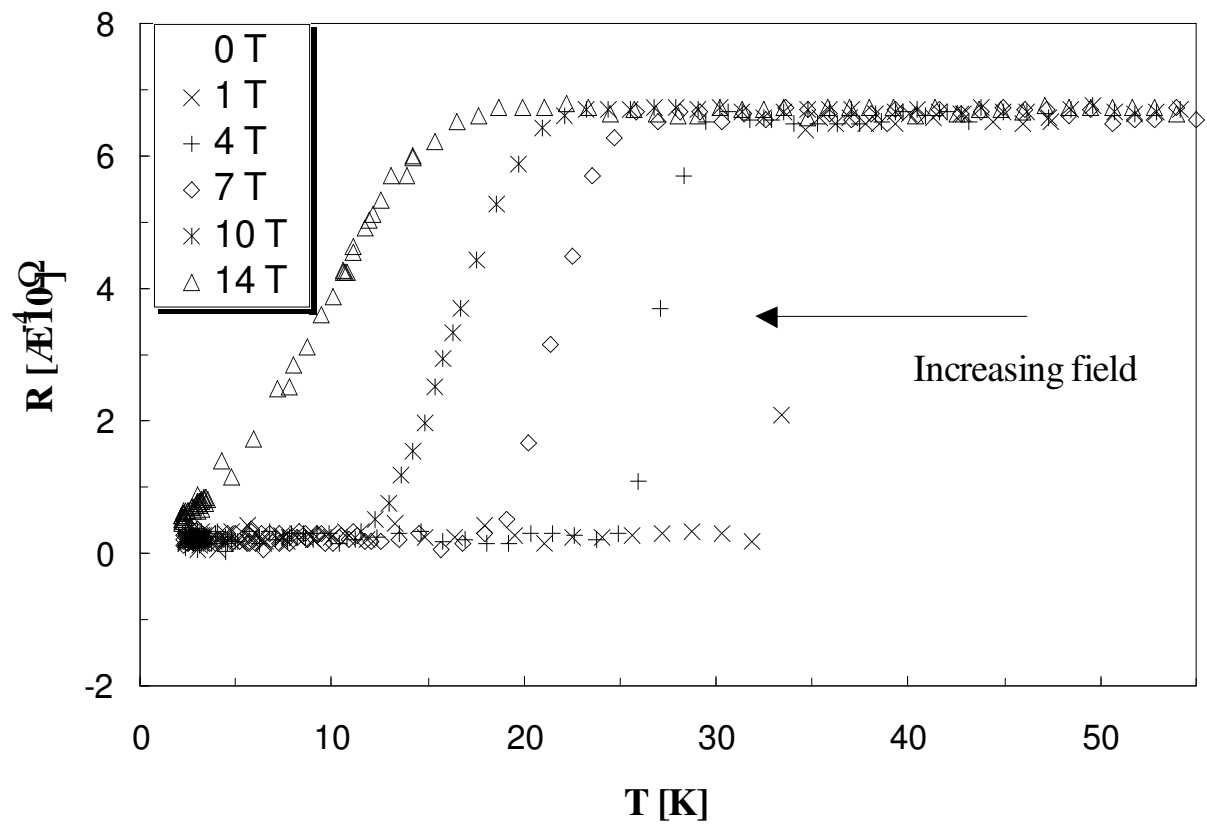
\end{document}